\documentclass[journal=jpccck,manuscript=article,layout=traditional,email=True]{achemso}
\usepackage[version=3]{mhchem}
\usepackage{achemso}
\usepackage{color}

\newcommand{\onlinecite}[1]{\hspace{-1 ex} \nocite{#1}\citenum{#1}}
\author{Rafi Chesler}
\affiliation{Department of Physics, University of Arizona}
\author{Souratosh Khan}
\affiliation{School of Information, University of Arizona
Tucson, AZ 85721}
\author{Sumit Mazumdar}
\affiliation{Department of Physics, University of Arizona}
\alsoaffiliation{Department of Chemistry and Biochemistry, University of Arizona}
\alsoaffiliation{College of Optical Sciences, University of Arizona}
\email{mazumdar@email.arizona.edu}
\date{\today}

\title{Wavefunction-Based Analysis of Dynamics Versus Yield of Free Triplets in Intramolecular Singlet Fission}

\begin{document}

\begin{abstract}
Experiments in several intramolecular singlet fission materials have indicated that the triplet-triplet spin biexciton has a much longer lifetime than believed until recently, opening up loss mechanisms that can annihilate the biexciton prior to its dissociation to free triplets. We have performed many-body calculations of excited state wavefunctions of hypothetical phenylene-linked anthracene molecules to better understand linker-dependent behavior of dimers of larger acenes being investigated as potential singlet fission candidates. The calculations reveal unanticipated features that we show carry over to the real covalently-linked pentacene dimers. Dissociation of the correlated triplet-triplet spin biexciton and free triplet generation may be difficult in acene dimers where the formation of the triplet-triplet spin biexciton is truly ultrafast. Conversely, relatively slower biexciton formation may indicate smaller spin biexciton binding energy and greater yield of free triplets. Currently available experimental results appear to support this conclusion. 
Whether or not the two distinct behaviors are consequences of distinct
mechanisms of triplet-triplet generation from the optical singlet is an interesting theoretical question.

\begin{tocentry}
\includegraphics[width=3.5in, height=1.5in]{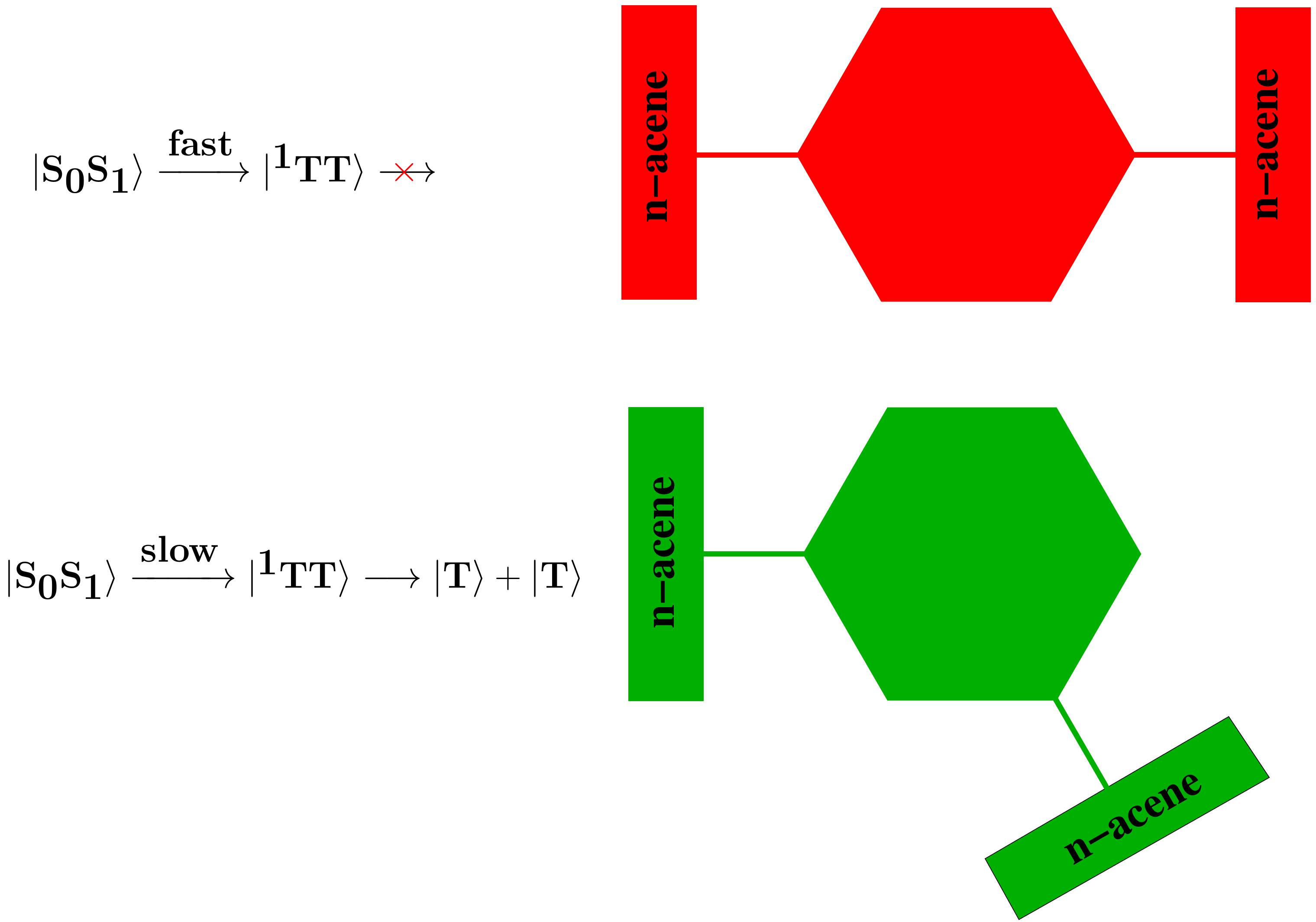}
\label{toc}
\end{tocentry}

\end{abstract}

\maketitle

\vskip 1pc

\section{Introduction}
Singlet-fission (SF), a spin-allowed multichromophore process in which two spin-triplet excitons are generated from a single optically allowed spin-singlet exciton, is of strong current interest \cite{Smith10a,Smith13a}, as successful implementation of SF can double the performance of organic solar cells \cite{Lee09a,Rao10a,Congreve13a}. While initial research focused mostly on molecular systems that exhibit intermolecular SF (xSF), interest has more recently been extended to covalently linked dimers and oligomers of chromophore molecules that exhibit intramolecular SF (iSF) \cite{Sanders15a,Zirzlmeier15a,Lukman15a,Busby15a,Fuemmeler16a,Sakuma16a,Sanders16a,Sanders16c,Korovina16a,Liu15b,Margulies16a,Korovina18b,Pun19a}. The work reported here is primarily concerned with iSF in covalently-liked dimers of chromophore molecules, although the broad conclusion we arrive at applies also to xSF.

In what follows, we use the ket notation to denote quantum-mechanical eigenstates, which are understood to be superpositions of many-electron configurations. The absolute ground state is written as $|S_0S_0\rangle$, the lowest optical singlet and triplet as  $|S_0S_1\rangle$ and  $|S_0T_1\rangle$, respectively (with the understanding that the actual symmetry-adapted eigenstates include configurations in which the molecules in their ground and excited states are switched). It is broadly accepted that SF is a two-step process in which $|S_0S_1\rangle$ first undergoes internal conversion to a bound triplet-triplet spin-singlet spin-biexciton, $|^1TT\rangle$. Free triplets are generated from dissociation of $|^1TT\rangle$. Based on transient absorption (TA) studies it was believed until recently that free triplets $T_1$ were being generated in ultrafast time scales (from hundreds of fs to ps). It is now realized that the earlier experiments were mostly detecting not free triplets but the bound $|^1TT\rangle$ \cite{Weiss16a,Yong17a,Stern17a,Tayebjee17a,Trinh17a,Folie18a,Pensack18a,Miyata19a,Musser19a}, whose lifetime can extend from ns to $\mu$s. Direct decay of the $|^1TT\rangle$ into $|S_0S_0\rangle$ or triplet-triplet annihilation are both possible over such a long period, indicating that actual implementation of SF in photovoltaics will require underdstanding both steps of SF equally well \cite{Pensack18a}. Theoretical research on SF has until now mostly focused on step 1, as determining accurate description of $|^1TT\rangle$ continues to be a formidable problem within traditional quantum chemical approaches.

The $|^1TT\rangle$ is a double excitation from the Hartree-Fock (HF) ground state in the molecular orbital (MO) representation and for molecules with more than 8-10 electrons is beyond the scope of traditional quantum-chemical approaches. Determination of the precise energy and wavefunction of the $|^1TT\rangle$ requires including configuration interaction (CI) with at least the dominant quadruple excitations from the HF ground state. We have theoretically investigated the $|^1TT\rangle$ using such high order CI in pentacene crystals \cite{Khan17a}, as well as covalently linked bipentacenes (BPn) \cite{Khan17b}, pentacene-tetracene heterodimers (PTn) \cite{Khan18a}, and the $para$- and $meta$-bisethynylpentacenylbenzene dimers (hereafter $p$- and $m$-Pc2) \cite{Khan20a} (see Fig. S1 of Supporting Information for the chemical structures of the four dimers). We showed that with the exception of $m$-Pc2 in all other cases $|^1TT\rangle$ should exhibit excited state absorptions (ESAs) in the near infrared (NIR) and mid-IR, {\it in addition to ESAs in the visible expected from free triplets.} These theoretical predictions have been confirmed from TA measurements in pentacene crystals \cite{Stuart19a,Zhou20a,Grieco18a} and BPn \cite{Trinh17a,Miyata19a} (see also discussion of $p$-Pc2 in reference \onlinecite{Khan20a}), confirming the long lifetimes of $|^1TT\rangle$ in these systems. In $m$-Pc2, TA from $|^1TT\rangle$ in the IR is neither expected theoretically \cite{Khan20a} nor seen experimentally \cite{Zirzlmeier15a}. Free triplet generation in $m$-Pc2 is likely, in agreement with the theoretically predicted weak binding energy of the $|^1TT\rangle$ \cite{Khan20a,Abraham17a}.

The ESAs from $|^1TT\rangle$ in the IR are due to inter-monomer charge-transfer (CT) excitations \cite{Khan17a,Khan17b,Khan18a,Khan20a}. CT from $|S_0S_1\rangle$, real or virtual, drives the transition to $|^1TT\rangle$ \cite{Smith10a,Smith13a}, while the extent of CT contribution to $|^1TT\rangle$ determines the binding energy E$_b$ of the $|^1TT\rangle$ (defined as E$_b = 2 \times E|S_0T_1\rangle-E|^1TT\rangle$, where $E|\cdots\rangle$ is the energy of the state being referred to) \cite{Khan20a}. Since the magnitude of E$_b$ determines the rate of $|^1TT\rangle$ dissociation, it follows that full many-body understanding of the role of CT in $|S_0S_1\rangle$ and $|^1TT\rangle$ will give qualitative understanding of the relationships between steps 1 and 2 of SF, if any. This is the goal of the present work.

Among iSF chromophores, the first step of SF is extremely rapid in $p$-Pc2 \cite{Zirzlmeier15a}, but E$_b$ here is moderate to large \cite{Khan20a}, while $m$-Pc2 is almost at the other extreme with slow internal conversion \cite{Zirzlmeier15a} but small E$_b$ \cite{Khan20a,Abraham17a}. We expect therefore that further study of $para$- and $meta$-linked chromophores will give deeper ``global'' insight into the two steps in SF. We have therefore performed calculations of many-body wavefunctions of hypothetical phenylene-linked para- and meta-bianthracenes (hereafter $p-2$ and $m-2$) shown in Fig.~1. We recognize at the outset that the compounds of Fig.~1 do not exist currently, and also that the real molecules will be nonplanar, unlike the planar dimer molecules investigated earlier. The advantage of calculations based on these model planar molecules is that the active space we are able to retain (24 MOs, see Fig.~2) constitutes a very large fraction of the total number of $\pi$-orbitals (38). The correlated-electron wavefunctions that we obtain, including that of $|^1TT\rangle$, are therefore near-exact within our model. We show that these near-exact wavefunctions, in spite of the small sizes of the model systems, provide very clear understanding of not only the experimental differences between $p$-Pc2 and $m$-Pc2, but that the physical understanding reached can be extended to other iSF systems (BPn and PTn) that we had investigated before. Based on our analysis of wavefunctions we argue that the yield of free triplets may be greater in SF systems with relatively slower dynamics, and vice versa. Although rapid annihilation of the $|^1TT\rangle$ characterizes the bulk of the iSF materials, free triplets in bulk amounts have been observed in a limited number of materials. We discuss these experimental observations in the Discussions and Conclusion section within the context of our theory.

\vskip 0.5pc

\begin{figure}
\includegraphics[width=3.5in]{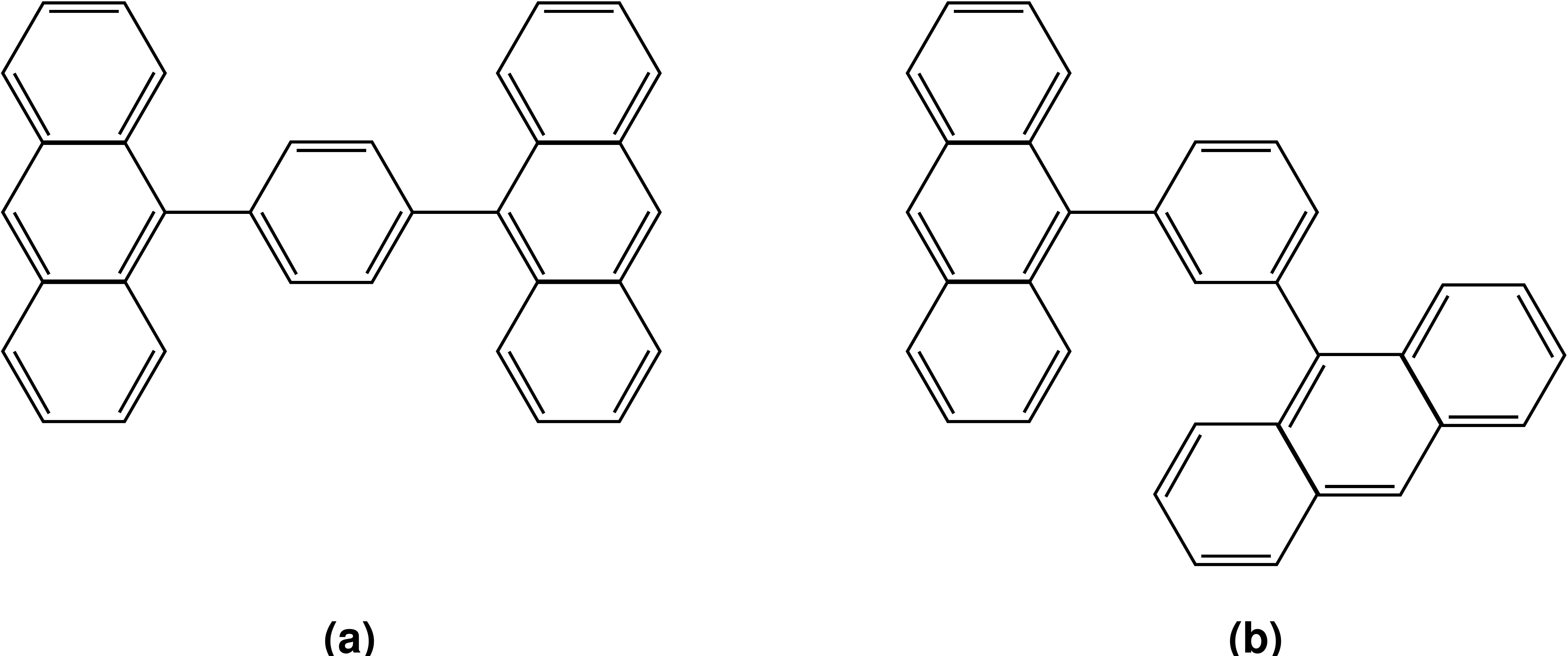}
\caption{\it The hypothetical (a) para and (b) meta-bianthracene molecules.} 
\label{molcules}
\end{figure}
\vskip 0.5pc

\section{Theoretical Model, Computational Methods and Parametrization}

\noindent \underbar {Model Hamiltonian.} Our work is based on the $\pi$-electron only Pariser-Parr-Pople (PPP) Hamiltonian \cite{Pariser53a,Pople53a},
\begin{equation}
H=\sum_{\langle ij \rangle,\sigma}t_{ij}(c_{i\sigma}^{\dagger}c_{j\sigma}+c_{j\sigma}^\dagger c_{i\sigma}) + U\sum_{i}n_{i\uparrow}n_{i\downarrow} + \sum_{i<j} V_{ij} (n_i-1)(n_j-1)
\label{PPP_Ham}
\end{equation}
where $c^{\dagger}_{i\sigma}$ creates an electron with spin $\sigma$ on the $p_z$ orbital of carbon (C) atom $i$, $n_{i\sigma} = \sum c^{\dagger}_{i\sigma} c_{i\sigma}$ is the number of electrons with spin $\sigma$ on atom $i$, and $n_i=\sum_{\sigma}n_{i\sigma}$ is the total number of electrons on the atom. We retain electronic hoppings $t_{ij}$ only between nearest neighbor carbon (C) atoms $i$ and $j$. $U$ is the Coulomb repulsion between two electrons occupying the $p_z$ orbital of the same C-atom, and $V_{ij}$ is long range Coulomb interaction. 

Our choice of the semiempirical PPP model over first principles approaches is based on the need to include electron correlations to high order. The apparent disadvantages of semiempirical models are well known: (i) the parameters entering the calculations can appear to be arbitrary; (ii) geometries of the molecules and bond lengths usually have to be assumed; and (iii) it is difficult to include electron-molecular vibration interactions accurately enough to account for differences in excited state and ground state geometries. Nevertheless, the semiempirical model remains the only choice when dynamic correlations have to be taken into account to high order, in order to determine $^1$TT energy and wavefunction {\it at the same level of accuracy} as S$_1$ and T$_1$ energies and wavefunctions. As was shown by Tavan and Schulten many years ago in the context of linear polyenes, even though the $^1$TT (2$^1$A$_g$) is predominantly a two electron-two hole (2e-2h) excitation, CI including upto 4e-4h quadruples is essential for obtaining the correct relative energy of this state vis a vis the S$_1$ (1$^1$B$_u$) state in polyenes with 10 or more $\pi$-electrons \cite{Tavan79a}. This is because, (a) 2e-2h excitations are coupled by the many-electron interactions in Eq.~\ref{PPP_Ham} to the 4e-4h interactions, and (b) even as the relative weight of any individual 4e-4h configuration to the $^1$TT is small, the {\it number} of 4e-4h configurations that contribute to the $^1$TT increases rapidly with the number of $\pi$-electrons. The PPP model allows us to to perform multiple reference singles and doubles CI (MRSDCI) calculations \cite{Tavan87a}, which retains the dominant 4e-4h excitations for target eigenstates. Note also that the consequences of electron-vibration interactions can often be guessed from the calculations of bond orders in the excited states within the PPP model \cite{Goli16a}.

\noindent \underbar{MRSDCI.} Unlike in our previous work, where we calculated ESAs from target states \cite{Khan17a,Khan17b,Khan18a,Khan20a}, here we are interested only in the energies and wavefunctions of the latter, and ground state absorption. The computations are thus simpler and more accurate \cite{Tavan87a}. For each target excited state we choose a trial set of appropriate 1e-1h and 2e-2h excitations from the single particle self-consistent MO ``ground state'' (see below) that best describe the target state. A double CI calculation is performed using these trial configurations, thereby generating the $N_{ref}$ reference configurations that make significant contributions to the target state. At the next step a higher order CI calculation that includes up to double excitations from the $N_{ref}$ reference configurations is performed. This generates both 3e-3h and 4e-4h excitations, as well as new 1e-1h and 2e-2h excitations coupled to these higher order excitations by the many-electron interactions in Eq.~\ref{PPP_Ham}. The new 1e-1h and 2e-2h excitations along with the original 1e-1h and 2e-2h configurations form the next set of $N_{ref}$ reference configurations and the procedure is repeated until convergence is reached. Convergence here is defined by the inclusion of the most dominant ne-nh (n = 1,2,3,4) configurations in the normalized wavefunctions of targeted excited states, with co-efficient $\geq$ 0.04. Typically, convergence requires $N_{ref}$ of the order of a few hundred and N$_{total}$ several times 10$^6$.

\noindent \underbar{Molecular exciton basis.} Our goal is physical interpretation of all eigenstates and to this end it is desirable to distinguish not only between predominantly 1e-1h versus predominantly 2e-2h excitations, but also between predominantly
intraunit versus interunit excitations (where by unit we refer to the individual anthracene monomers and the phenylene linker in Fig.~1). Within the molecular exciton basis, the one-electron MO basis consists of simply the products of the MOs on the three individual units (see Fig.~2). While similar approach has also been used by others for calculations of SF \cite{Smith13a,Japahuge18a,Casanova18a,Tempelaar17a,Fuemmeler16a}, these other calculations usually retain only the highest occupied and lowest unoccupied MO (HOMO and LUMO) of the individual units and a handful (usually between 5-10) of many-electron configurations. Our basis space exceeds 10$^6$ for each targeted wavefunction.

We rewrite Hamiltonian~\ref{PPP_Ham} as $H_{intra} + H_{inter}$, where $H_{intra}$ describes each individual unit, and $H_{inter}$ consists of interunit $t_{ij}$ and components of $V_{ij}$ where atoms $i$ and $j$ belong to different units. $H_{intra}$ are first solved separately for each unit within the Hartree-Fock (HF) approach, and the solution simply consists of products of HF MOs of the individual units (as in Fig.~2). The basis functions for the complete Hamiltonian $H_{intra} + H_{inter}$ are many-electron configurations with all possible electron occupancies of the localized MOs of Fig.~2 including upto 4e-4h excitations from the ``ground state'' configuration that has all bonding MOs in Fig.~2 fully occupied. The basis space for the different target states we have investigated are in all cases several million within the MRSDCI approach (see Supporting Information Section C). In what follows we show for each correlated wavefunction only the most dominant configurations that have the largest relative weights, along with their normalized CI coefficients.
\begin{figure}
\includegraphics[height=3.5in]{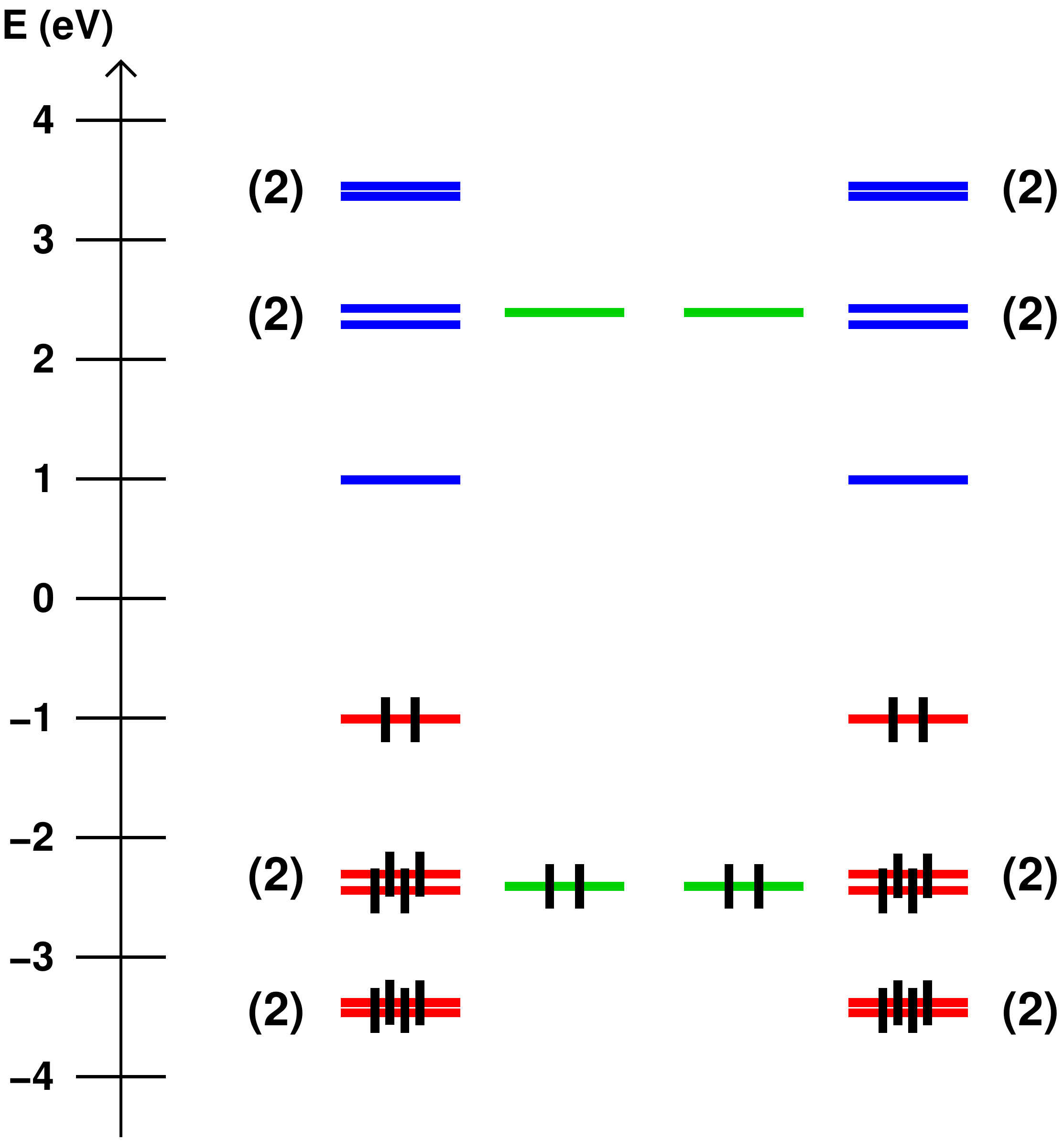}
\caption{\it Huckel MOs localized on the individual anthracene molecules and the phenylene linker in the bianthracenes 
that constitute the active space for the PPP-MRSDCI calculations. Red and blue horizontal lines correspond to bonding and antibonding MOs of the
anthracenes, while the green horizontal lines correspond to the doubly degenerate HOMOs and LUMOs of benzene.  
The numbers in parentheses 
give the number of close-lying MOs.  The $y$-axis
gives the energy with our hopping integrals.}
\label{MOs}
\end{figure}
Because of the complex natures of the correlated wavefunctions in bianthracene to which the linker MOs make signifcant contributions, we explain here our notation by listing explicit examples of a select set of the diagrammatic many-electron basis configurations in Fig.~3. In all cases we show the HOMO and LUMO of the individual anthracene MOs and their occupancies. {\it MOs excluded in any diagram have the same occupancies as in Fig.~2,} viz., the excluded bonding (antibonding) MOs are completely occupied (unoccupied). The absence of the linker phenylene MOs in any configuration therefore indicates that they are ``inactive'' in that particular configuration. Configurations in which the phenylene linker MOs do participate in the excitations in bianthracene are also included in Fig.~3. Only one each of the doubly degenerate HOMO and LUMO of the benzene molecule is delocalized over the entire molecule \cite{Salem66a}, and only these delocalized MOs participate in interunit charge-transfer. Hence in the diagrammatic representations of configurations involving charge-transfer between anthracene and phenylene only a single bonding or antibonding benzene MO is included. Note that participation by the localized benzene MO in configurations with {\it intraphenylene} excitations are however not precluded. Lines linking singly excited MOs in Fig.~3 are spin-singlet bonds; these lines are replaced by arrows to describe spin-triplet bonds. We have not drawn bonds linking MOs in the case of double excitations. It is implied in the total spin $S=0$ case that each such diagrammatic double excitation includes all three S$_z$ components (+1-1, 00 and -1+1) where S$_z$ is the z-component of the total spin.

\vskip 1pc

\begin{figure}
\includegraphics[width=3.5in]{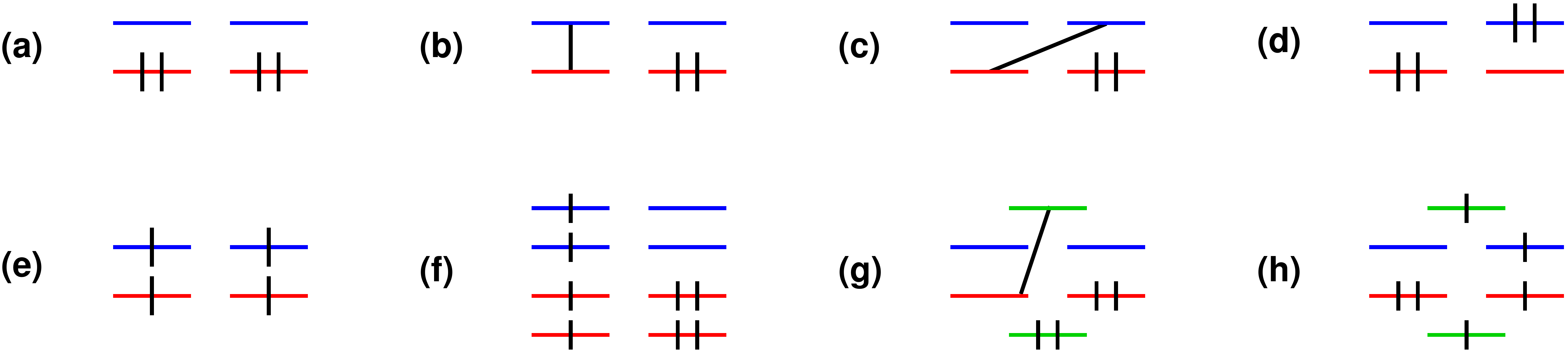}
\vskip 1pc
\caption{\it Sample $S=0$ diagrammatic basis functions.
Bonding (antibonding) MOs absent in any diagram are completely filled (unoccupied). Lines connecting MOs represent singlet bonds. (a) The ``ground state'' configuration {\bf G}. (b) One of two Frenkel excitations {\bf FE}. (c) One of two configurations {\bf CT} with charge-transfer between anthracene units. (d) One of two configurations {\bf D} with doubly excited anthracene monomers. 
(e) Double excitation {\bf TT} with 1e-1h excitations on both monomers; {\bf TT} can be both spin singlet ($S=0$) and spin quintet ($S=2$). (f) Double excitations involving multiple bonding and antibonding MOs on anthracene units. (g) Configurations with charge-transfer between anthracene monomer and the phenylene linker; only the delocalized benzene MOs are shown (see text).
(h) Double excitation involving anthracene and phenylene linker.}
\label{basis}
\end{figure}
   
\noindent \underbar {Parametrization.} We have chosen peripheral and internal bond lengths for the anthracene monomers to be (1.40 $\mathring{\textrm{A}}$) and (1.46 $\mathring{\textrm{A}}$) \cite{Khan18a}, respectively, and the corresponding $t_{ij}$ as $-$2.4 and $-$2.2 eV, respectively \cite{Ducasse82a}. Both $p-2$ and $m-2$ are assumed to be planar, with bond lengths 1.40 $\mathring{\textrm{A}}$ and $t_{ij}=-2.4$ eV for the C-C bonds internal to the phenylene linker and bond length 1.46 $\mathring{\textrm{A}}$ and $-2.2$ eV, respectively, for the interunit bonds between the anthracene monomers and the phenylene linker. We use the screened Ohno parameterization for the long range Coulomb repulsion, $V_{ij}=U/\kappa\sqrt{1+0.6117 R_{ij}^2}$, where $R_{ij}$ is the distance in $\mathring{\textrm{A}}$ between C-atoms $i$ and $j$ and $\kappa$ is an effective dielectric constant \cite{Chandross97a}. The Coulomb parameter $U$ (7.7 eV) and the dielectiric constant $\kappa$ (1.3) are chosen by quantitative fitting of the anthracene monomer singlet and triplet excitation energies (see Supporting Information).

\vskip 0.5pc

\noindent \section {Results and Analysis}
In the following we present the results of our computations on $p-2$ and $m-2$, comparing these against our previous computational results for pentacene dimers \cite{Khan17b,Khan18a,Khan20a} wherever appropriate. We find that in spite of the greater contribution of the linker MOs to the wavefunctions of $p-2$ and $m-2$, theory finds that the fundamental photophysics are the same for our model molecules and the larger systems.

\vskip 0.5pc

\noindent \underbar {The Ground States of $p-2$ and $m-2$.} We have given in Fig.~4 the most dominant components of the
correlated ground states of $p-2$ and $m-2$ for $U=7.7$ eV, $\kappa=1.3$. Our most important conclusions are, (i) the relative weight of {\bf G}, as measured by the square of its CI coefficient, in the near-exact $|S_0S_0\rangle$ eigenstates of Figs. 4(a) and (b) is only about 65\%, (ii) $|S_0S_0\rangle$ has small but nonzero contribution from {\bf D}, but vanishingly small contribution from {\bf TT}, and (iii) somewhat surprisingly, the ground state wavefunctions of these differently linked dimer molecules are practically identical. {\it The strong admixing of configurations in $|S_0S_0\rangle$ (in particular the relatively small contribution by {\bf G}), which is separated from $|S_0S_1\rangle$ and $|^1TT\rangle$ by large energy gaps, clearly indicates the futility of trying to guess the behavior of the $|^1TT\rangle$, which is proximate to other eigenstates, from the character of {\bf TT} alone.}

\begin{figure}
\includegraphics[width=6.5in]{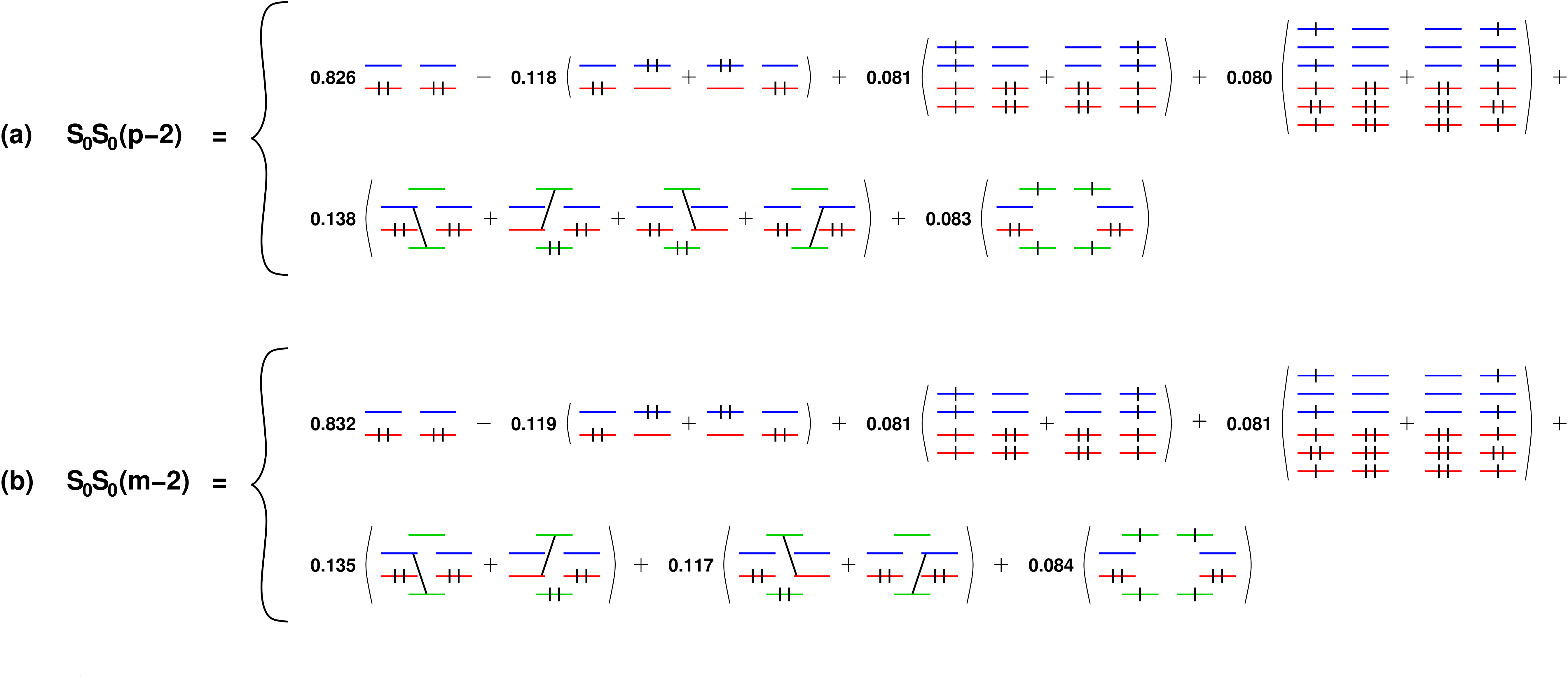}
\caption{\it The near-exact normalized ground state wavefunctions (a) $p-2$, (b) $m-2$. Note the relatively strong contributions by {\bf D} and other intra-monomer multiple excitations, but absence of {\bf TT}.}
\label{ground-state}
\end{figure}

\vskip 0.5pc

\noindent \underbar{Excited state energies.} In Table I we have given the excited state energies of $|S_0S_1\rangle$, $|S_0T_1\rangle$, and the spin-singlet and spin-quintet triplet-triplets $|^1TT\rangle$ and $|^5TT\rangle$ for $p-2$ and $m-2$. We have also included the spin-gap $\Delta_S$ = E($|^5TT\rangle$)-E($|^1TT\rangle$), and E$_b$. The significant differences between the energies of the optical singlet state and of the triplet exciton in the monomer versus the dimer indicate greater delocalization in $p-2$ and $m-2$, compared to BPn \cite{Khan17b}, PTn \cite{Khan18a} and $p$- and $m$-Pc2 \cite{Khan20a}; the monomer versus dimer energy differences are vanishingly small in dimers of the larger acenes. The larger delocalization in $p-2$ and $m-2$ is, however, between anthracene monomers and the phenylene linker and not between the anthracene monomers themselves (see below). The very small $\Delta_S$ and E$_b$ in Table I for $m-2$ are completely in agreement with what was previously obtained in $m$-Pc2 \cite{Khan20a}, which is surprising, given the apparently delocalized nature of $|^1TT\rangle$ here. This is the first sign that delocalization into the linker molecule does not affect the relationships between the essential singlet, triplet and triplet-triplet states and conclusions based on calculations on these model molecules can be extended to other dimers.

\vskip 1pc

\noindent {\it Table 1. Excited state energies (in eV) of $p-2$ and $m-2$, along with their spin gaps $\Delta_S$ and binding energy E$_b$. $\Delta_S$ $>$ E$_b$ in $p-2$ is a finite size effect \cite{Khan20a}}

\vskip 2.0pc

\begin{tabular}{|l c c c c c c|}
\hline Compound & $S_0S_1$ & $S_0T_1$ &
$^1TT$ & $^5TT$ & $\Delta_S$ & $E_b$ \\ \hline
$p-2$ & 2.81 & 1.68 & 2.97 & 3.51 & 0.53 & 0.38 \\ 
$m-2$ & 3.03 & 1.73 & 3.43 & 3.42 & -0.01 & 0.03 \\ \hline
\end{tabular}

\vskip 2.0pc

\noindent \underbar{Singlet and triplet exciton wavefunctions.} In Fig. 5 we have shown the wavefunctions for $|S_0S_1\rangle$ in $p-2$ and $m-2$, respectively. The optical states are remarkably different in the two dimer molecules, unlike the ground state wavefunctions. While {\bf CT} configurations make perceptible contribution to $|S_0S_1\rangle$ in $p-2$, their contribution to the same state in $m-2$ is vanishingly small. This same difference was previously perceived between the $|S_0S_1\rangle$ wavefunctions of $p$- and $m$-Pc2 \cite{Khan20a}. Other authors have made similar observations based on the symmetry characters of the MOs in the $meta$-compounds \cite{Ito16a}. The absence of inter-anthracene charge-transfer in $m-2$ occurs here in spite of e-h delocalization into the linker phenylene unit.

The $|S_0T_1\rangle$ states of $p-2$ and $m-2$, shown in Fig.~6, are different from one another in the same way as the spin singlet states. The anthracene monomers are therefore nearly completely decoupled in $|S_0S_1\rangle$ and $|S_0T_1\rangle$ states of $m-2$. The contribution by {\bf CT}, relative to the {\bf FE} configuration is significantly smaller in $|S_0T_1\rangle$ than $|S_0S_1\rangle$ of $p-2$, indicating that the lowest triplet is more localized than the optical singlet, a feature generally true
with $\pi$-conjugated systems within correlated-electron models.

\begin{figure}
\includegraphics[width=6.5in]{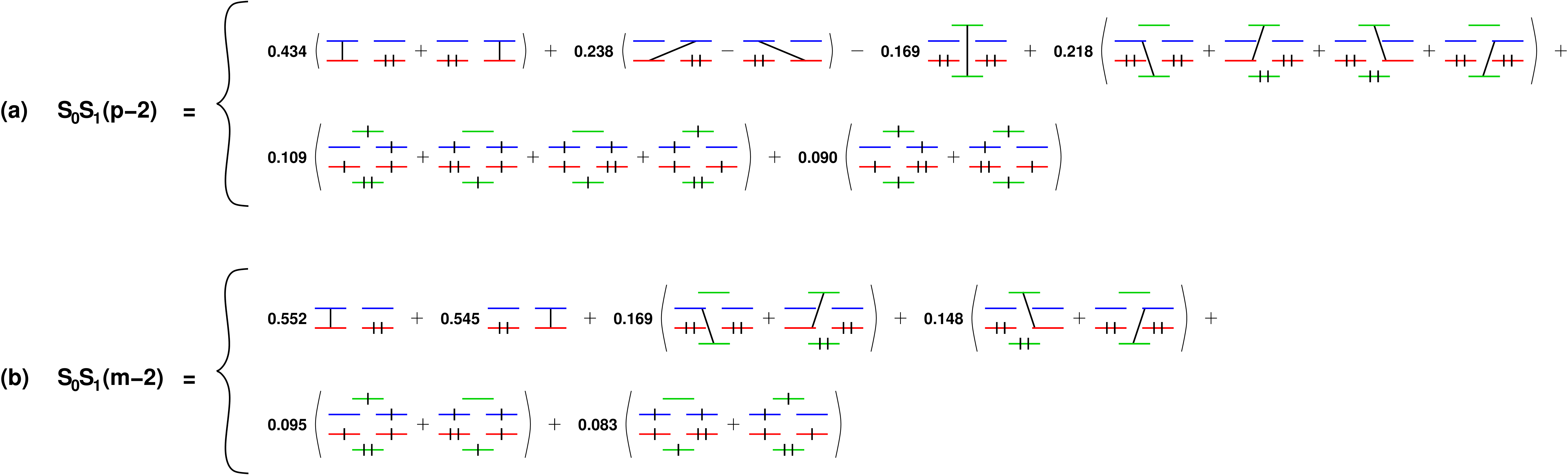}
\caption{\it Optical spin-singlet wavefunctions of (a) $p-2$, (b) $m-2$.}
\label{optical-state}
\end{figure}

\vskip 0.5pc

\begin{figure}
\includegraphics[width=6.5in]{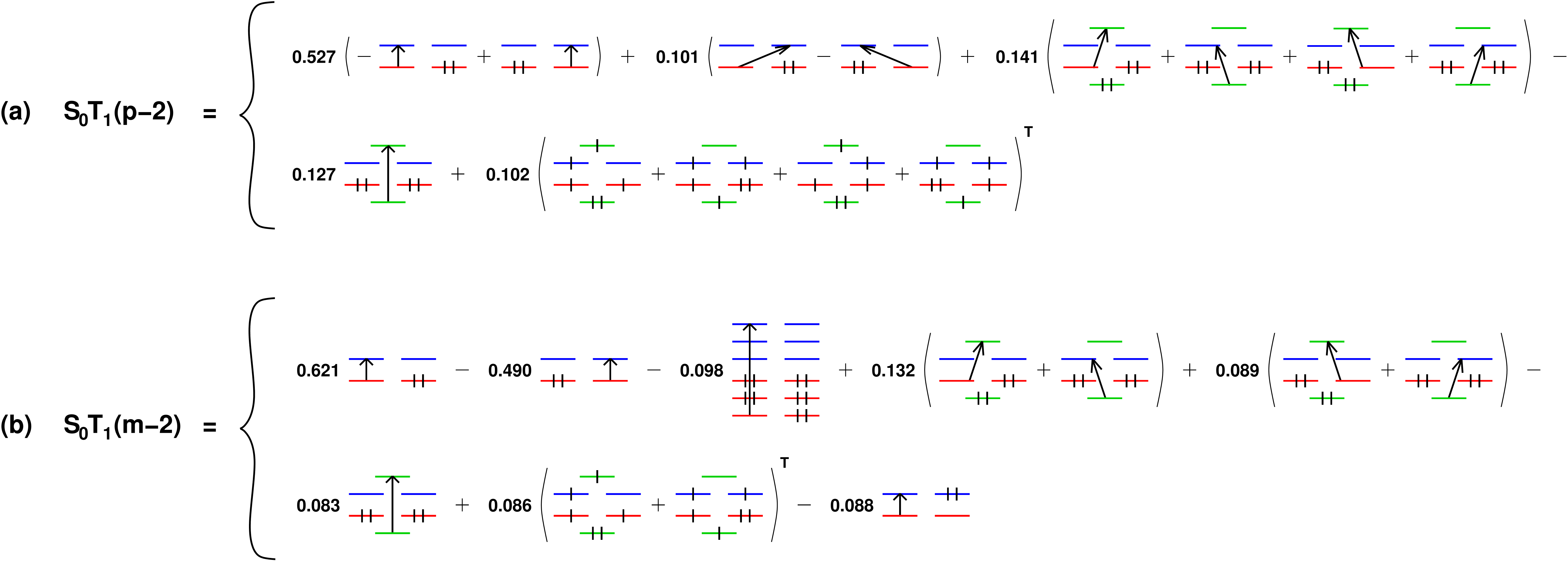}
\caption{\it Lowest spin triplet wavefunctions of (a) $p-2$, (b) $m-2$. The arrows designate spin-triplet bonds, inclusive of all three S$_z$ components. The labels ``T'' on the double excitations indicate that the configurations have total spin $S=1$.}
\label{triplet}
\end{figure}

\vskip 0.5pc

\noindent \underbar{The triplet-triplet spin-biexcitons.} 
In Fig. 7 we have shown the  $|^1TT\rangle$ and $|^5TT\rangle$ wavefunctions for $p-2$. Not surprisingly, {\bf TT} makes the largest contribution to both wavefunctions, although the contributions are very different to the two spin states. Importantly, the contribution of {\bf D} is nearly zero. As was found previously for stacks of polyenes \cite{Aryanpour15a}, there is nearly complete decoupling between double excitation within a single monomer versus double 1e-1h excitations on different acene monomers. The spin-singlet and quintet wavefunctions are very different in character. The strong contributions by {\bf G} and {\bf CT} to $|^1TT\rangle$ and their absence in the $|^5TT\rangle$ are both expected, because of the overall spin $S=0$ character of these configurations. The large $\Delta_S$ and E$_b$ of $p-2$ (see Table I) are also due to the strong contributions by {\bf G} and {\bf CT} to $|^1TT\rangle$. With increased size of the acene monomer, the relative weights of configurations in which the linker MOs are involved decreases and that of {\bf TT} increases \cite{Khan20a}.

\begin{figure}
\includegraphics[width=6.5in]{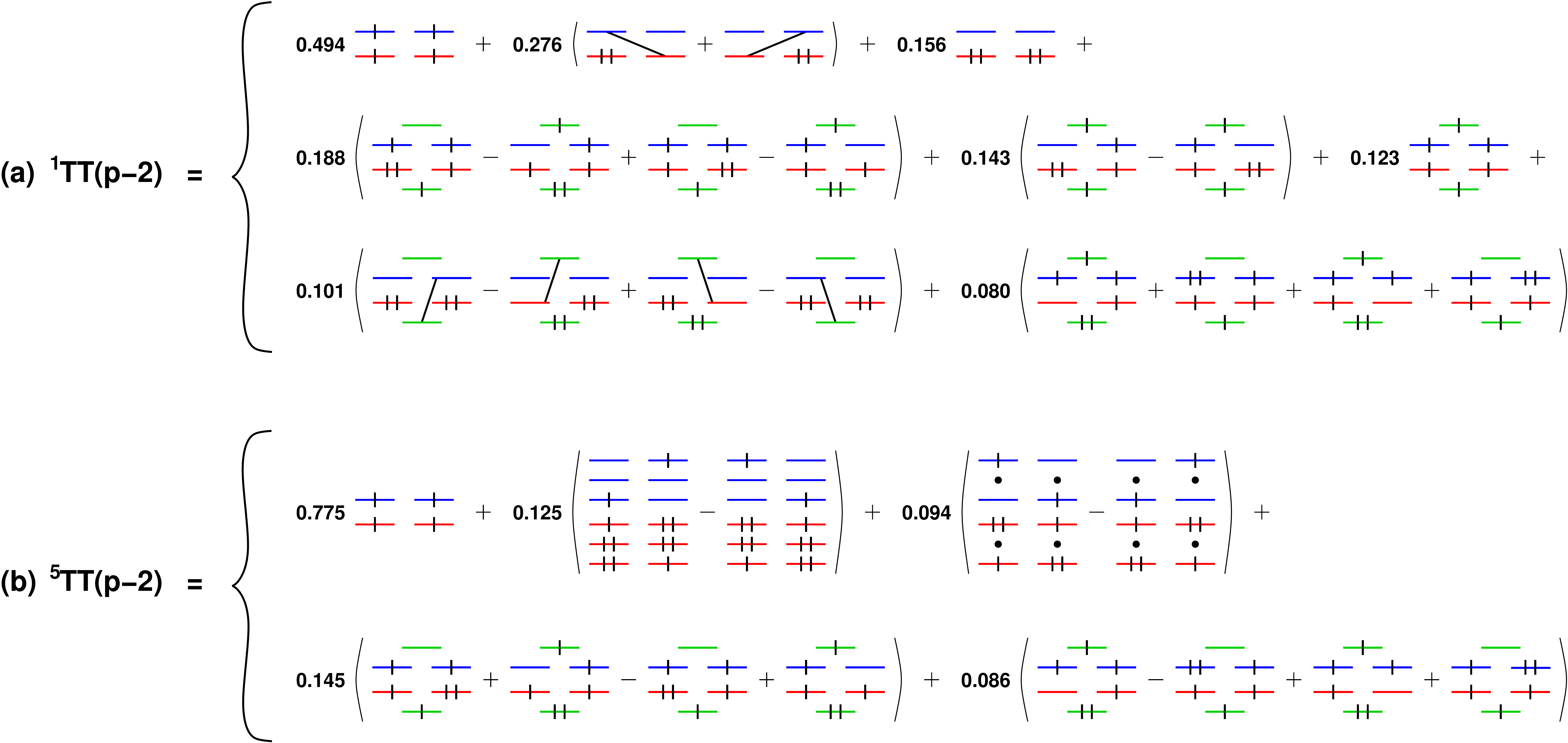}
\caption{\it Normalized wavefunctions (a) $|^1TT\rangle$ and (b) $|^5TT\rangle$ of $p-2$. The dots represent multiple intervening MOs.
Admixing with {\bf G} and {\bf CT} gives the nonzero binding energy of the $S=0$ spin-biexciton $|^1TT\rangle$.}
\label{TTp-2}
\end{figure}

Fig.~8 shows the  $|^1TT\rangle$ and $|^5TT\rangle$ wavefunctions for $m-2$. The absence of {\bf CT} in $|^1TT\rangle$ is expected, based on the absence of the same in $|S_0S_1\rangle$ and $|S_0T_1\rangle$. Note, however, that this occurs in spite of the triplet-triplet states occurring {\it above} $|S_0S_1\rangle$, 
indicating that the Heisenberg spin-Hamiltonian description of eigenstates \cite{Abraham17a} is not a requirement for absence of {\bf CT}. More importantly, we note the rather remarkable result that {\it the $|^1TT\rangle$ and $|^5TT\rangle$ wavefunctions in $m-2$ are practically identical}: the relative weights of each individual configuration in the two normalized wavefunctions are the same, which is of course a consequence and signature of the very weak coupling between the anthracene monomers. Returning to Fig.~7, we also note that $|^5TT\rangle$ in $p-2$ is closely related to these eigenstates.

Given the results of Fig.~8 we returned to our work in reference \onlinecite{Khan20a} and determined the accurate triplet-triplet wavefunctions for $p$-Pc2 and $m$-Pc2 to high order (see Supporting Information Figs.~S2 and S3). We have found that the identical natures of $|^1TT\rangle$ and $|^5TT\rangle$ wavefunctions is true even for the real dimer $m$-Pc2. The spin-biexcitons occur eneregetically below $|S_0S_1\rangle$ in $m$-Pc2 and the wavefunctions are almost entirely superpositions of {\bf TT} and higher energy 2e-2h excitations on the anthracene monomers. The linker MOs are largely inactive. As with $m-2$ the relative weights of each individual configuration is the same in both spin states of $m$-Pc2. This is in contrast to the $|^1TT\rangle$ and $|^5TT\rangle$ wavefunctions of $p$-Pc2, which are again different from one another.

\begin{figure}
\includegraphics[width=6.5in]{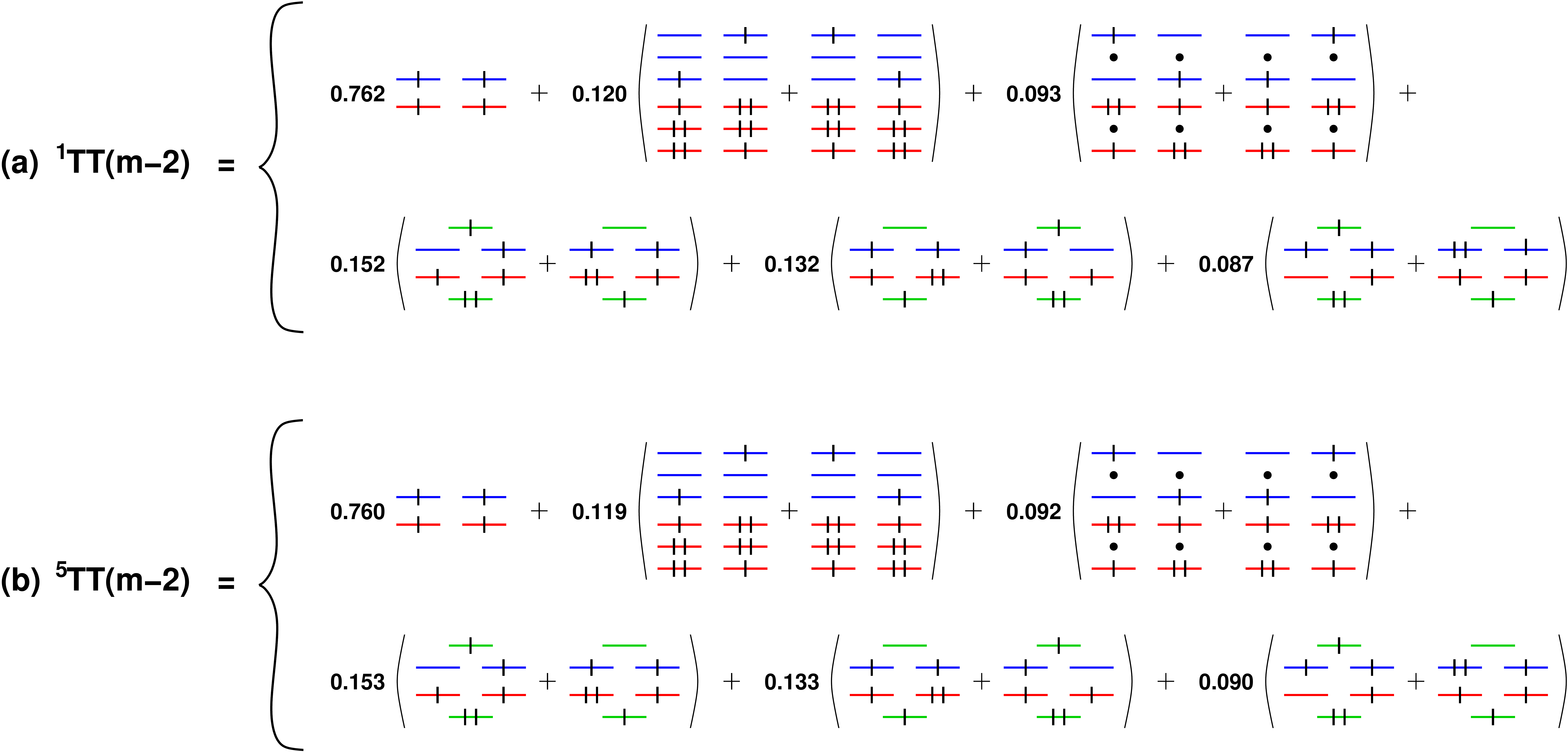}
\caption{\it Normalized wavefunctions (a) $|^1TT\rangle$ and (b) $|^5TT\rangle$ of $m-2$. The coefficients of individual configurations
are the same in (a) and (b), indicating that the wavefunctions are completely identical.}
\label{TTm-2}
\end{figure}
 
\noindent \section{Discussions and Conclusion}

The very large active space we have used for correlated-electron calculations on $p-2$ and $m-2$ gives us confidence about the accuracy of the wavefunctions we obtain. In spite of greater linker MO contributions to the wavefunctions in these smaller molecules we find that they give the correct information for the realistic dimers. The strong admixing of {\bf CT} configurations
in both $|S_0S_1\rangle$ and $|^1TT\rangle$ in $p-2$ and in the pentacene dimers BPn, PTn and $p$-Pc2 investigated previously indicates that SF in these systems is charge-transfer mediated. The relative weight of {\bf CT} in $|S_0S_1\rangle$ in these cases directly controls the coupling between $|S_0S_1\rangle$-$|^1TT\rangle$ and hence the time $\tau_{TT}$ in which the $|^1TT\rangle$ is formed \cite{Zeng14a,Aryanpour15a,Monahan15a,Feng16a,Alvertis19a,Chen20a} (we ignore here the requirement that $|S_0S_1\rangle$ to $|^1TT\rangle$ transition must involve a symmetry-breaking interaction). To the best of our knowledge, there is currently no explicit discussion of the extent of {\bf CT} mixing in $|^1TT\rangle$ and the consequence of such mixing. Our calculations for $p-2$ indicate that the relative weight of {\bf CT} is significantly larger in $|^1TT\rangle$ than in $|S_0T_1\rangle$. The greater {\bf CT} content of $|^1TT\rangle$ contributes to its lower energy relative to $2\times$ E($|S_0T_1\rangle$). We further note that {\bf CT} contributions to energetically proximate $|S_0S_1\rangle$ and $|^1TT\rangle$ are comparable (see Figs.~5(a) and 7(a)). This is simply a consequence of energy proximity E($|S_0S_1\rangle$) $\simeq$ E($|^1TT\rangle$) within the correlated-electron Hamiltonian, which in turn is a requirement for effcient SF. Taken together, it follows that short $\tau_{TT}$ and moderate to large E$_b$ go together. We note in this context that in $m-2$ we find {\bf CT} content to be extremely small in $|S_0S_1\rangle$, $|^1TT\rangle$ and $|S_0T_1\rangle$. This would indicate both longer $\tau_{TT}$ and smaller E$_b$. We have gone through the previously obtained wavefunctions for BPn, PTn, $p$-Pc2 and $m$-Pc2, and we have found in all cases {\bf CT} contributions to individual $|S_0S_1\rangle$ and $|^1TT\rangle$ are comparable, indicating that in all cases smaller $\tau_{TT}$ is accompanied by larger E$_b$, and vice versa. Fast $|^1TT\rangle$ generation may therefore indicate that it does not further dissociate onto free triplets. In Table II we have given the calculated energies of the optical singlet and the triplet-triplet, along with the experimentally determined $\tau_{TT}$ and our calculated E$_b$ for BP0 and BP1, PT0 and PT1,
and $p$-Pc2 and $m$-Pc2. A qualitative one-to-one correspondence between {\it experimental} $\tau_{TT}^{-1}$ and calculated E$_b$ is seen (the calculated E$_b$ for BP1 is an outlier; the large error in the calculation is an indication that even an active space of 24 MOs is not sufficient for giving precise E$_b$ here).

\vskip 1pc

\noindent {\it Table II. Calculated energies of the optical singlet and the triplet-triplet, the experimental time-constant of formation of the triplet-triplet (in ps) and the calculated binding energy of the latter for covalently-linked pentacene dimers. In both PT0 and PT1 there are two optical singlet states. The calculated quantities are from references \onlinecite{Khan17b} (BPn), \onlinecite{Khan18a} (PTn) and \onlinecite{Khan20a} ($p$-Pc2 and $m$-Pc2), respectively. $a$, $b$ and $c$ correspond to experimental references \onlinecite{Sanders15a,Sanders16a,Zirzlmeier15a}, respectively.} 

\vskip 2.0pc

\begin{tabular}{|p{0.15\textwidth} p{0.15\textwidth} p{0.15\textwidth} p{0.15\textwidth} p{0.15\textwidth}|}

\hline Compound & $E|S_0S_1\rangle$ (eV) & $E|^1TT\rangle$ (eV) & $\tau_{TT}$ (ps) & E$_b$ (eV) \\[4pt] \hline
BP0 & 1.91 & 1.75 & 0.76$^a$ & 0.08 \\
BP1 & 1.91 & 1.75 & 20$^a$ & 0.13 \\
PT0 & 1.90, 2.17 & 1.97 & 0.83$^b$ & 0.09 \\
PT1 & 1.91, 2.19 & 2.00 &18.3$^b$ & 0.06 \\
$p$-Pc2 & 1.81 & 1.71 & 2.7 $\pm$ 1.0$^c$ & 0.08 \\
$m$-Pc2 & 1.85 & 1.74 & 80$^c$ & 0.009 \\ \hline

\end{tabular}

\vskip 1pc

As pointed out in the Introduction, in most experiments with iSF compounds free triplets are not the end products in bulk amounts. Experiments that do detect free triplets appear to agree with our theoretically arrived at conclusion about the correspondence between short $\tau_{TT}$ and small yield of free triplets, as we now point out.

(i) In recent experiments on ortho- and meta-linked phenylene-bridged pentacene dimers that are structurally different from the compounds we have investigated (see Fig. S4 of Supporting Information for the chemical structures) Sakai {\it et al.} have found that the rate constants for $|^1TT\rangle$ formation are 1.2$\times$10$^{11}$ s$^{-1}$ in the ortho- compound and
2.1$\times$10$^9$ s$^{-1}$ in the meta-compound, respectively \cite{Sakai18a}. The smaller rate constant in the meta-compound is clearly due to the weak contribution of {\bf CT} to $|S_0$S$_1\rangle$, as found in our calculations for $m-2$ and $m$-Pc2. The quantum yields of $|^1TT\rangle$ in the two compounds are comparable. The authors performed time-resolved electron spin resonance on both systems, and concluded that the quantum yield of free triplets in the ortho-dimer with $\tau_{TT}$ two orders of magnitude faster is only 20\%, in contrast to quantum yield greater than 100\% in the meta-dimer \cite{Sakai18a}. The comparable quantum yields of $|^1TT\rangle$ in the two compounds suggests that the loss in the ortho-dimer occurs in the second step of SF, presumably due to the larger E$_b$ of $|^1TT\rangle$ here. Vanishing E$_b$ in the meta-compound would be expected from our calculations.

(ii) Copious yields of free triplets have been also obtained in a {\it nonconjugated} pentacene dimer \cite{Basel17a} (see Fig. S4 of Supporting Information for the chemical structure). Based on the absence of conjugation configuration mixing between {\bf FE} and {\bf CT}, as well as between {$^1$TT} and {\bf CT} are expected to be very weak here. Not surprisingly, the experimentally determined rate constant for $|^1TT\rangle$ formation (2.4$\times$10$^9$ s$^{-1}$) is significantly smaller than in most conjugated dimers, and is comparable to the meta-linked pentacene dimer mentioned above \cite{Sakai18a}. 

(iii) While our calculations are for iSF dimers, the conclusion regarding the correspondence between short $\tau_{TT}$ and larger E$_b$ is arrived at from noting the nearly comparable {\bf CT} contents in $|S_0S_1\rangle$ and $|^1TT\rangle$, and should be equally true for xSF materials, where also the energies of these states are comparable. Pensack {\it et al.} have found precisely such relationship in their measurements on amorphous pentacene nanoparticles with three different sidegroups. The authors find that although triplet-triplet quantum yield is independent of the size of the sidegroup, $\tau_{TT}$ increases with the size of the sidegroup (as expected) and free triplet yield is highest in the nanoparticle with the longest $\tau_{TT}$ \cite{Pensack18a}. 

In conclusion, we calculated near-exact correlated wavefunctions of the optical singlet and triplet excitons, and the spin-singlet and quintet triplet-triplet biexcitons of hypothetical bianthracenes, $p-2$ and $m-2$, within the PPP Hamiltonian for realistic Coulomb correlations. The calculations reveal that with the exception of the ground state and $|^5TT\rangle$, there is strong admixing of {\bf CT} configurations in all eigenstates of $p-2$. Conversely, {\bf CT} plays very weak role in $m-2$, in which the $|^1TT\rangle$ and $|^5TT\rangle$ wavefunctions are identical. The identical characters of $|^1TT\rangle$ and $|^5TT\rangle$ are true for the real dimer $m$-Pc2. Based on the available experimental results \cite{Zirzlmeier15a} and our previous and current theoretical work we suggest time-resolved electron spin resonance measurments in $m$-Pc2 to determine whether free triplets are being generated here too. Theoretical research on SF until now has largely concluded that it is mediated by charge-transfer \cite{Smith13a,Japahuge18a,Casanova18a,Tempelaar17a,Zeng14a,Monahan15a}. Given the nature of the iSF compounds in which free triplets have been observed \cite{Sakai18a,Basel17a} it is conceivable that charge-transfer plays a very weak role in these. Whether or not the direct mechanism of iSF \cite{Fuemmeler16a} is more relevant in compounds that have shown free triplet formation is an interesting theoretical question and a topic of future research.

\begin{suppinfo}
TIPS-acene dimers BP0, BP1, PT0, PT1, $p-Pc2$ and $m-Pc2$, linked by phenylene spacer groups, experimental and calculatd singlet and triplet energies of anthracene, Hamiltonian matrix dimensions for ground state, optical singlet, triplet, and triplet-triplet states for $p-2$ and $m-2$, $^1$TT and $^5$TT
wavefunctions of pentacene dimers $p-Pc2$ and $m-Pc2$, TIPS-pentacene dimers whose dynamics and free triplet yields have been investigated recently.
\end{suppinfo}

\section{Acknowledgments.} S.M. is thankful to Professor Joseph Michl, University of Colorado  for the invitations to the
summer workshops on singlet fission that allowed him to become aware of the advances being made in the field by research groups from all over the world.
This research was supported by grant NSF-CHE-1764152. 
 
\bibliography{proposal-chem-new}

\end{document}